\documentclass{article}
\usepackage{etoolbox}
\newtoggle{arxiv}
\toggletrue{arxiv} 

\usepackage{xcolor}
\usepackage{amsmath}
\usepackage{url}

\definecolor{myblue}{rgb}{0.21,0.49,0.74}
\iftoggle{arxiv}{
    \usepackage[bookmarks=false, colorlinks, allcolors=myblue,pdfborder={0 0 0},pdfborderstyle={/S/U/W 0}]{hyperref}

}{
    \usepackage[bookmarks=false,pdfauthor={\authorname},pdfsubject={\pdfsubject},hidelinks]{hyperref}
}

\usepackage[T1]{fontenc}
\usepackage[utf8]{inputenc}

\iftoggle{arxiv}{
    \usepackage{ismir}
}{
    \usepackage[submission]{ismir} 
}

\usepackage{graphicx}
\usepackage{color}
\graphicspath{{figures/}}
\usepackage{mdframed}
\usepackage[most]{tcolorbox}
\usepackage{subcaption}
\usepackage[inline]{enumitem}
\usepackage{linguex}

\newcommand\blfootnote[1]{%
  \begingroup
  \renewcommand\thefootnote{}\footnote{#1}%
  \addtocounter{footnote}{-1}%
  \endgroup
}

\usepackage{placeins}
\usepackage{url}
\usepackage{booktabs}
\usepackage{amsmath, amsfonts}
\usepackage{overpic}
\usepackage{amssymb}
\usepackage{appendix}
\usepackage{todonotes}
\usepackage{leftindex}
\usepackage{algorithm}
\usepackage{algorithmic}
\usepackage{multirow}
\usepackage{scalerel}
\usepackage{tikz}
\usetikzlibrary{3d}  
\usepackage{tikz-3dplot}
\usepackage{algorithmic}
\usepackage{dsfont}
\usepackage{colortbl}
\usepackage{tabularx}

\usepackage{wrapfig}
\usepackage{arydshln}

\usepackage{cleveref}

\newcommand{\diffmean}[0]{\texttt{DiffMean}}
\newcommand{\musicgen}[0]{\texttt{MusicGen}}
\newcommand{\melody}[0]{\texttt{MusicGen-Melody}}

\usepackage{tikz}
\usetikzlibrary{positioning,calc,arrows.meta,shapes}

\multauthor
  {Simone Facchiano$^{1, \star}$ \hspace{1cm} Giorgio Strano$^{1, \star}$ \hspace{1cm} Donato Crisostomi$^{1}$}
  {{\bf Irene Tallini$^{1}$ \hspace{1cm} Tommaso Mencattini$^{2}$ \hspace{1cm} Fabio Galasso$^{1}$  \hspace{1cm} \hspace{1cm} Emanuele Rodolà$^{1}$ }\\
  $^{1}$Sapienza University of Rome, $^{2}$École Polytechnique Fédérale de Lausanne\\
  {\tt\small \{facchiano, strano\}@di.uniroma1.it}
  }

\def\authorname{F. Author, S. Author, and T. Author}

\title{Activation Patching for Interpretable Steering in Music Generation}

\makeatletter
\def\adl@drawiv#1#2#3{%
  \hskip.5\tabcolsep
  \xleaders#3{#2.5\@tempdimb #1{1}#2.5\@tempdimb}%
  #2\z@ plus1fil minus1fil\relax
  \hskip.5\tabcolsep}
\newcommand{\cdashlinelr}[1]{%
  \noalign{\vskip\aboverulesep
    \global\let\@dashdrawstore\adl@draw
    \global\let\adl@draw\adl@drawiv}
  \cdashline{#1}
  \noalign{\global\let\adl@draw\@dashdrawstore
    \vskip\belowrulesep}}
\makeatother

\newcommand{\vertical}[1]{\rotatebox[origin=c]{90}{#1}}

\begin{document} 

\maketitle

\begin{abstract}
    Understanding how large audio models represent music, and using that understanding to steer generation, is both challenging and underexplored. \blfootnote{$\star$ denotes equal contribution.}Inspired by mechanistic interpretability in language models, where direction vectors in transformer residual streams are key to model analysis and control, we investigate similar techniques in the audio domain. This paper presents the first study of latent direction vectors in large audio models and their use for continuous control of musical attributes in text-to-music generation. Focusing on binary concepts like tempo (fast vs. slow) and timbre (bright vs. dark), we compute {\em steering vectors} using the difference-in-means method on curated prompt sets. These vectors, scaled by a coefficient and injected into intermediate activations, allow fine-grained modulation of specific musical traits while preserving overall audio quality. We analyze the effect of steering strength, compare injection strategies, and identify layers with the greatest influence. Our findings highlight the promise of direction-based steering as a more mechanistic and interpretable approach to controllable music generation.
    \begin{center}
        \scriptsize
        \hspace{0.2cm}\raisebox{-0.2\height}{\includegraphics[width=1em,height=1em]{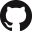}}\hfill \href{https://github.com/gladia-research-group/music-mechint}{\texttt{github.com/gladia-research-group/music-mechint}}

         \hspace{0.2cm}\raisebox{-0.2\height}{\includegraphics[width=1em,height=1em]{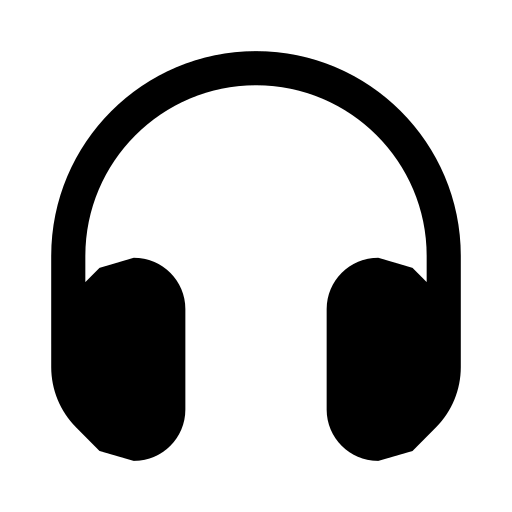}}\hfill \href{https://gladia-research-group.github.io/music-mechint-demo/}{\texttt{gladia-research-group.github.io/music-mechint-demo}}
    \end{center}
\end{abstract}

\section{Introduction}

\begin{figure}
    \centering
    \includegraphics[width=\linewidth]{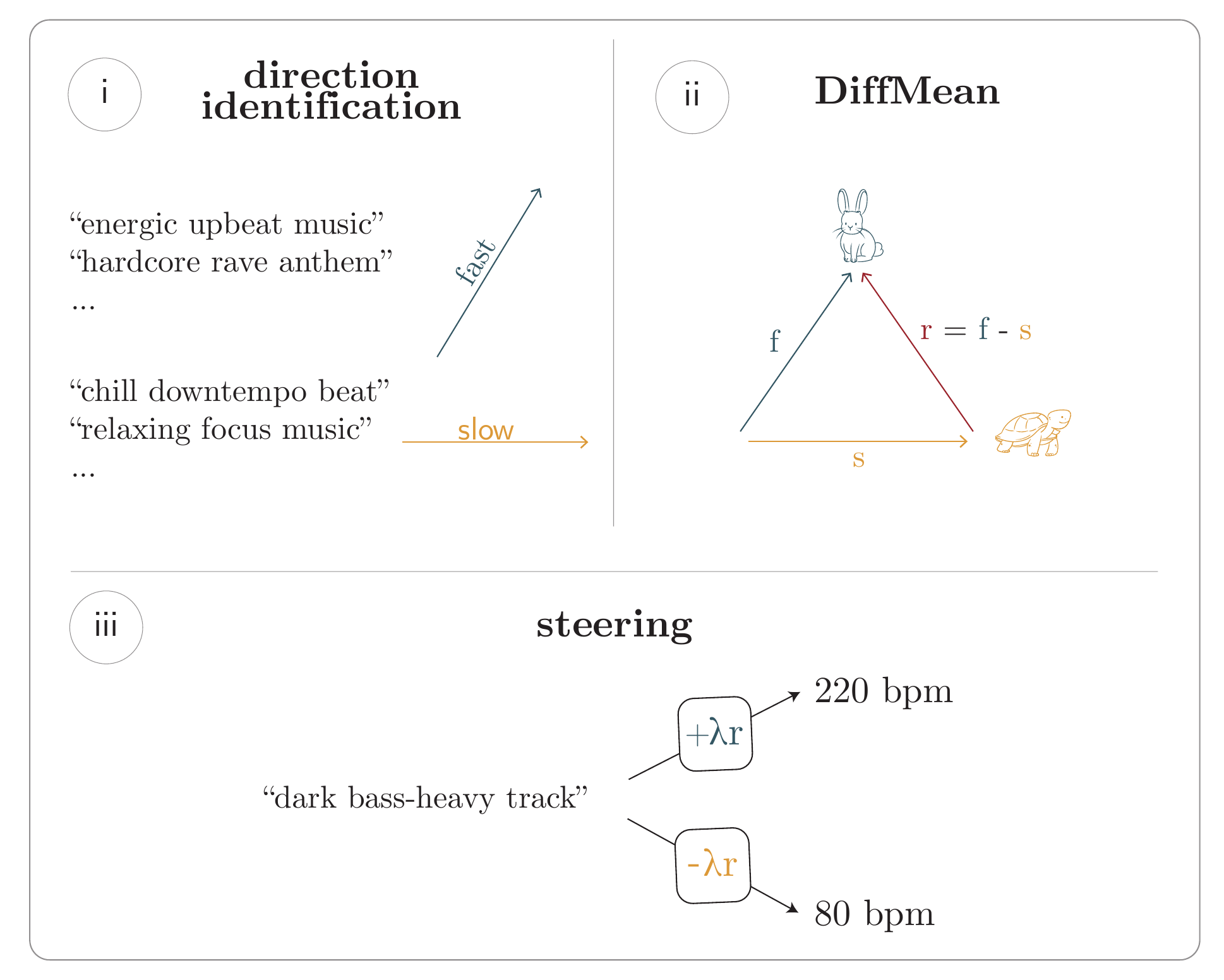}
    \caption{Steering pipeline. (i) We obtain relevant directions for each attribute (e.g. slow, fast) by averaging the embeddings of semantically-related prompts. (ii) We identify the direction controlling the attribute by taking the difference of these means. (iii) We steer the generation by either adding or removing the identified direction. }
    \label{fig:pipeline}
\end{figure}

Large language models (LLMs), originally developed for text, have extended seamlessly to audio \cite{agostinelli2023musiclm, musicgen, dhariwal2020jukeboxgenerativemodelmusic}, where they enable state-of-the-art music generation by predicting tokens from quantized neural codecs \cite{défossez2022highfidelityneuralaudio}. However, while mechanistic interpretability has provided powerful tools for understanding and controlling LLMs in natural language processing \cite{arditirefusal,kim2018interpretability,voita2019bottom,subramani2022extractinglatentsteeringvectors,turner2024steeringlanguagemodelsactivation,konen2024stylevectorssteeringgenerative,tan2025analyzinggeneralizationreliabilitysteering,wu2025axbenchsteeringllmssimple}, its application to music models remains crucially underexplored. Unlike text, which follows a discrete, syntactic structure, \textit{music} is a continuous, high-dimensional signal where key attributes—such as tempo and timbre—lack a direct symbolic representation in model inputs and outputs.

This gap raises fundamental questions: Do text-to-music models encode high-level musical properties in a steerable way? Can intervention techniques developed for language models truly be adapted to control audio generation? Unlike text, music relies on highly granular tokenizations and much longer temporal structures, making it challenging to align short textual prompts with rich, continuous musical outputs. These differences complicate the direct application of existing intervention methods. Consequently, answering these questions is critical both for advancing interpretability in multimodal models and for practical applications, such as enabling fine-grained stylistic control over music generation.

In this paper, we introduce for the first time a mechanistic interpretability framework applied to music generation by investigating whether \textit{activation-based steering} \cite{subramani2022extractinglatentsteeringvectors,turner2024steeringlanguagemodelsactivation, konen2024stylevectorssteeringgenerative, marksgeometry, belrose2023diffinmeans, panickssery2023steering} can manipulate specific musical attributes. While this method can generally be applied to any pair of contrasting attributes, in this study we focus on two fundamental binary concepts: tempo (fast vs.\ slow) and timbre (bright vs.\ dark). Using the \diffmean{} method \cite{marksgeometry,belrose2023diffinmeans, panickssery2023steering}, we extract latent directions that differentiate contrastive sets of prompts and evaluate whether injecting these steering vectors into the model’s intermediate activations can systematically shift the generated audio along the intended dimension.

To do this, we curate prompt sets that elicit contrasting musical outputs and compute the corresponding activation differences across transformer layers. These direction vectors are then injected {\em at inference time} to assess their impact on generated music (see \Cref{fig:pipeline}). We conduct an extensive study of factor affecting the \diffmean{} method, comparing injection strategies, identifying key layers, and evaluating how scaling and dataset dimensionality of the steering prompts influence outcomes. In total, we generate over 20,000 samples per attribute pair, providing a robust foundation for current and future analysis.

Our findings are as follows:
\begin{itemize}
    \item Steering vectors effectively modulate musical attributes, showing that activation-based interventions can influence high-level generation properties.
    \item A mid-range block of transformer layers is most responsible for encoding and controlling tempo and brightness, suggesting a structured representation of musical features.
    \item Steering strength scales with a coefficient $\lambda$, where moderate values yield smooth control, while extreme values can introduce audio artifacts.
\end{itemize}

The rest of this paper is organized as follows: \Cref{sec:method} introduces the \diffmean{} steering approach for audio models. \Cref{sec:experiments} details our experimental setup, including prompt design, evaluation metrics, and layer-wise analysis. Results are presented in \Cref{sec:results}, followed by a discussion of interpretability implications in \Cref{sec:discussion}, along with limitations and future directions.

\section{Background and related work}

Our work builds on the growing literature on mechanistic interpretability. We begin by reviewing key techniques, with a focus on activation engineering.

\subsection{Mechanistic Interpretability}

Mechanistic interpretability aims to understand and control the computational mechanisms underlying deep neural networks \cite{arditirefusal,kim2018interpretability,voita2019bottom,subramani2022extractinglatentsteeringvectors,turner2024steeringlanguagemodelsactivation,konen2024stylevectorssteeringgenerative,tan2025analyzinggeneralizationreliabilitysteering,wu2025axbenchsteeringllmssimple}. 
These can be broadly divided into observation and intervention approaches \cite{bereska2024mechanisticinterpretabilityaisafety}. The former analyze the internal workings of a neural network—such as activations, weights, or learned representations—without altering them, while intervention methods actively modify or manipulate these factors in order to gain deeper insights into how the model processes information.

One of the earliest observation methods is \textit{classifier probing} \cite{alain2016understanding, hewitt2019structural}. In classifier probing, researchers extract activations from one or more layers of a trained neural network and feed those activations into a simple classifier to determine whether specific properties (e.g., syntactic roles in a language model, semantic attributes, or other labeled features) can be predicted from the hidden representations. 

On the other hand, a key class of intervention techniques is \textit{activation engineering}, which modifies activations at inference time to influence the model’s output. A particularly effective strategy within this framework is the use of \textit{steering vectors} \cite{subramani2022extractinglatentsteeringvectors,turner2024steeringlanguagemodelsactivation,konen2024stylevectorssteeringgenerative,tan2025analyzinggeneralizationreliabilitysteering}—directions in activation space that correspond to specific attributes. By adding or removing these vectors from a model’s hidden states, researchers have successfully shifted output distributions along meaningful semantic dimensions.
Steering vectors can be computed in several ways. Early work optimized them through gradient-based methods \cite{subramani2022extractinglatentsteeringvectors}, while later approaches extracted them from contrastive activation differences between pairs of prompts \cite{turner2024steeringlanguagemodelsactivation}. More robust techniques take the \textit{mean activation} over two sets of prompts representing opposing concepts and compute the difference between them, a method known as \diffmean{} \cite{arditirefusal,marksgeometry,belrose2023diffinmeans,panickssery2023steering}. Crucially, \texttt{AxBench} \cite{wu2025axbenchsteeringllmssimple} found the latter to provide superior control over model behavior when compared to other interpretability-driven techniques  such as Sparse AutoEncoders (SAEs) \cite{cunningham2023sparse, attention_saes, makelov2024towards}. Building on these findings, we adopt the \diffmean{} method as our primary technique for computing steering vectors due to its robustness and performance.

\subsection{Language Modeling for Music Generation}
The concept of treating music generation as a language modeling task was introduced by Jukebox \cite{dhariwal2020jukeboxgenerativemodelmusic}, which represents audio as multi-scale discrete tokens produced by a residual VQ-VAE and generates them level by level using a hierarchical Transformer. Despite being able to produce consistent and long musical excerpts, it struggles with quantization artifacts.

With the advent of new residual quantized codecs, such as SoundStream \cite{soundstream2021zeghidour} and EnCodec \cite{défossez2022highfidelityneuralaudio}, the hierarchical paradigm of Jukebox was re-adopted in MusicLM \cite{agostinelli2023musiclm}, which couples semantic tokens (for high-level content) with SoundStream acoustic tokens (for fine-grained audio details), feeding them into a cascade of transformers. 
In contrast, MusicGen \cite{musicgen} follows a non-hierarchical approach, where a single Transformer autoregressively models EnCodec’s residual-quantized tokens. Due to its simplicity and state-of-the-art music generation quality, we focus on MusicGen.

MusicGen is available in four variants: small, medium, large, and \melody{}. The first three use text conditioning by encoding prompts with T5 \cite{raffel2023exploringlimitstransferlearning} and integrating the embeddings through cross-attention. In contrast, \melody{} directly prepends the text to the input sequence alongside a quantized melody representation (e.g., a chromagram extracted from reference audio). Since \melody{} operates within the same token domain for both conditioning and output, we use this variant throughout our experiments, applying it with empty melody conditioning to maintain consistency with the standard \musicgen{} setup. For brevity, we refer to this variant simply as \musicgen{} throughout the paper.

\subsection{Interpretability of Music Generative Models}
Interpretability research in the music domain is scarce and has this far been limited to probing. Probing has been applied to Jukebox and MusicGen to identify which layers can predict high-level tags (e.g., “a colorful happy violin song”), classify concepts such as genre, emotion, or key \cite{castellon2021codified, Koo2024understanding}, as well as to detect theoretical musical constructs like notes, scales, and intervals \cite{wei2024musicgenerationmodelsencode}. The results in \cite{wei2024musicgenerationmodelsencode} show that, consistently with our findings, probing accuracy for all tasks begins to increase around $20\%$ of the network depth, while those in \cite{Koo2024understanding} reveal that emotion detection capabilities emerge earlier (around $20\%$ of network depth) compared to key and genre detection, which appear later.

To the best of our knowledge, ours is the first work to explore whether the information encoded in the intermediate representations of a large audio model can be leveraged to steer generation. It is also the first study that investigates the role of direction vectors in large audio models.

\section{Methodology}
\label{sec:method}

\subsection{Steering Vector Computation}

\begin{figure}[t]
\centering
\includegraphics[width=0.7\linewidth]{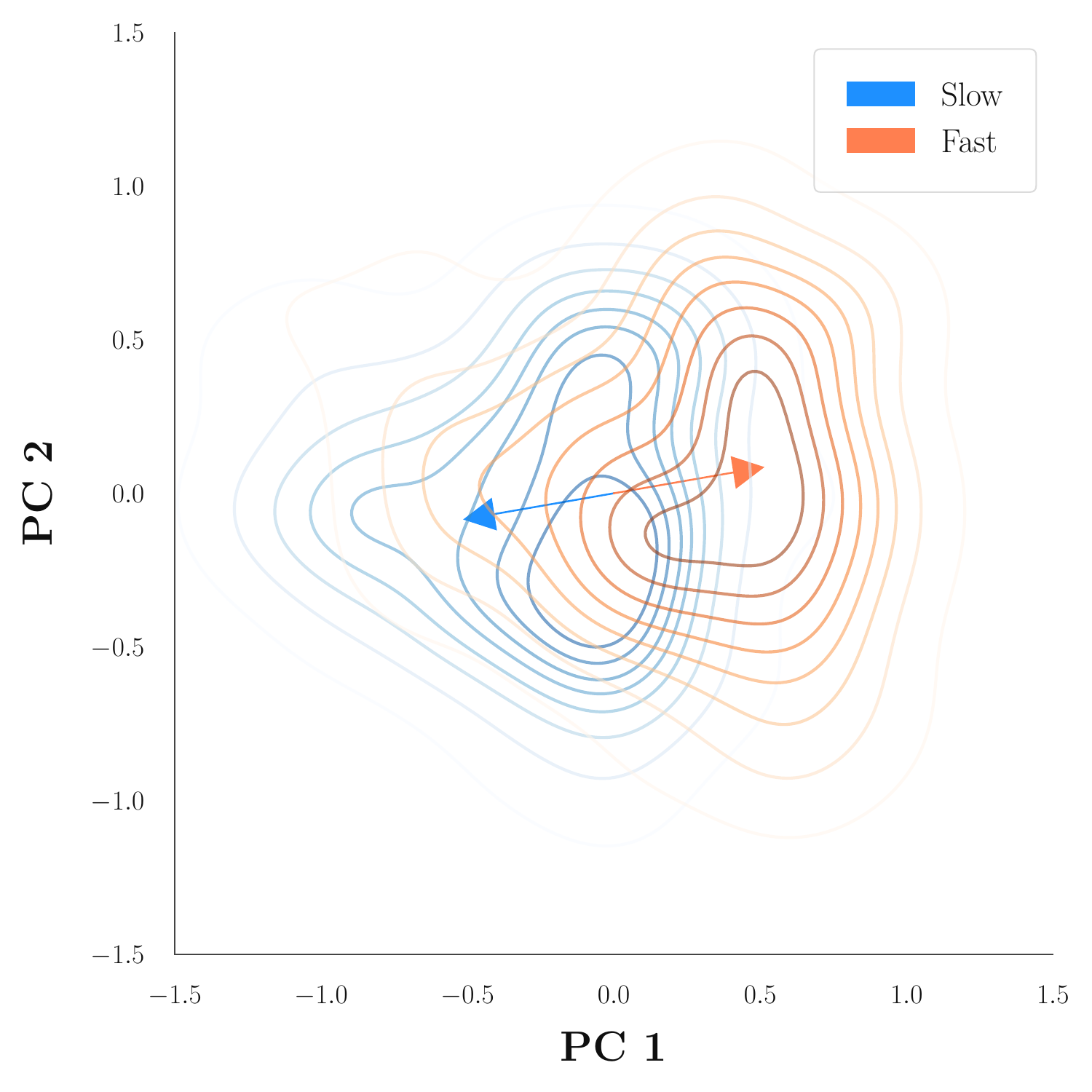}
\caption{2D Kernel Density Estimation of \musicgen{}'s activations, projected via PCA, at layer $14$ for pairs of prompts belonging to $S_{\text{Fast}}$ and $S_{\text{Slow}}$.}
\label{fig:PCA_directions}
\end{figure}

To investigate how \musicgen{} represents high-level musical concepts, we define a target attribute and create prompt sets that elicit contrasting model behaviors. For instance, to study tempo, we define $S_{\text{Fast}}$ and $S_{\text{Slow}}$, containing prompts that should generate fast and slow music, respectively. Similarly, we use $S_{\text{Bright}}$ and $S_{\text{Dark}}$ to analyze timbre.

We run the model over these prompts and extract hidden representations at every layer. To capture how the model encodes the prompt as a whole, we use the hidden state of the end-of-sequence (\texttt{EOS}) token, which summarizes the input. This results in two sets of activation vectors (one per concept) across all $L$ layers. 

At each layer, we average the activations within each set to obtain a pair of mean vectors. These mean activations define distinct {\em directions} in the model’s latent space that separate the two contrasting concepts. Their difference defines a \textit{steering vector} -- a direction in latent space that distinguishes the two concepts. \Cref{fig:PCA_directions} visualizes these directions in the case of fast vs. slow music.

Formally, let $M$ be a model with $L$ layers, and let $h^{(l)}(x)$ denote the hidden representation at layer $l$ when processing input $x$. Given two sets of prompts, $S_A$ and $S_B$, designed to elicit opposing attributes, we compute the mean activations for each set at every layer:
\[
\mu^{(l)}_A = \frac{1}{\lvert S_A\rvert} \sum_{x \in S_A} h^{(l)}(x), 
\quad
\mu^{(l)}_B = \frac{1}{\lvert S_B\rvert} \sum_{x \in S_B} h^{(l)}(x).
\]
The \diffmean{} vector at layer $l$ is then defined as:
\begin{equation}
\Delta^{(l)} = \mu^{(l)}_A - \mu^{(l)}_B \,.
\label{eq:steering_vector}
\end{equation}
This vector encodes the difference in how the model internally processes the two concepts. 

\begin{figure}[t]
    \centering
    \begin{tikzpicture}[
        scale=0.7,
        node distance=1.4cm,
        every node/.style={font=\small},
        align=center,
        >=stealth,
        thick,
        baseline=(current bounding box.north)
    ]
        \node at (0,1.0) {\textbf{All-to-All Strategy}};
        
        \node[draw, rectangle] (h1) {\(h^{(1)}\)};
        \node[draw, rectangle, right=1.4cm of h1] (delta1) {\(\Delta^{(1)}\)};
        \node[circle, draw, right=1.0cm of delta1] (plus1) {\(+\)};
        \node[draw, rectangle, right=1.0cm of plus1] (hp1) {\(h^{(1)}_{\text{steer}}\)};
        
        \draw[->] (h1) -- (delta1);
        \draw[->] (delta1) -- (plus1);
        \draw[->] (plus1) -- (hp1);

        \node[draw, rectangle, below of=h1, node distance=1.6cm] (h2) {\(h^{(L)}\)};
        \node[draw, rectangle, right=1.4cm of h2] (delta2) {\(\Delta^{(L)}\)};
        \node[circle, draw, right=1.0cm of delta2] (plus2) {\(+\)};
        \node[draw, rectangle, right=1.0cm of plus2] (hp2) {\(h^{(L)}_{\text{steer}}\)};
        
        \draw[->] (h2) -- (delta2);
        \draw[->] (delta2) -- (plus2);
        \draw[->] (plus2) -- (hp2);

        \node at ($(h1)!0.5!(h1.south)+(0,-0.8)$) {\(\vdots\)};
        \node at ($(hp1)!0.5!(hp1.south)+(0,-0.8)$) {\(\vdots\)};
    \end{tikzpicture}
    \begin{tikzpicture}[
        scale=0.7,
        node distance=1.4cm,
        every node/.style={font=\small},
        align=center,
        >=stealth,
        thick,
        baseline=(current bounding box.north)
    ]
        \node at (0,1.0) {\textbf{One-to-All Strategy}};
        
        \node[draw, rectangle] (h1) {\(h^{(1)}\)};
        \node[draw, rectangle, below right=0.3cm and 1.4cm of h1] (deltabest) {\(\Delta^{(l_{\text{best}})}\)};
        \node[circle, draw, right=3.35cm of h1] (plus1) {\(+\)};
        \node[draw, rectangle, right=1.0cm of plus1] (hp1) {\(h^{(1)}_{\text{steer}}\)};
        
        \draw[->] (h1) -- (deltabest);
        \draw[->] (deltabest) -- (plus1);
        \draw[->] (plus1) -- (hp1);

        \node[draw, rectangle, below of=h1, node distance=1.6cm] (h2) {\(h^{(L)}\)};
        \node[circle, draw, right=3.35cm of h2] (plus2) {\(+\)};
        \node[draw, rectangle, right=1.0cm of plus2] (hp2) {\(h^{(L)}_{\text{steer}}\)};

        \draw[->] (h2) -- (deltabest);
        \draw[->] (deltabest) -- (plus2);
        \draw[->] (plus2) -- (hp2);

        \node at ($(h1)!0.5!(h1.south)+(0,-0.8)$) {\(\vdots\)};
        \node at ($(hp1)!0.5!(hp1.south)+(0,-0.8)$) {\(\vdots\)};
    \end{tikzpicture}

    \caption{
    \textbf{(Top)} In \emph{All-to-All}, each layer \(l\) receives its own steering vector \(\Delta^{l}\). 
    \textbf{(Bottom)} In \emph{One-to-All} instead, a single direction \(\Delta^{(l_{\text{best}})}\) is injected into every layer.
    }
    \label{fig:injection_strategies}
\end{figure}
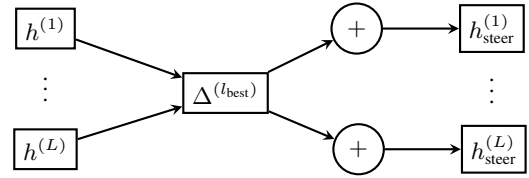 

\subsection{Steering the Generation}
To steer the model’s behavior at inference time, we modify its hidden states by injecting $\Delta^{(l)}$ scaled by a coefficient $\lambda$:
\begin{equation}
    h^{(l)}_{\text{steer}} \leftarrow h^{(l)} + \lambda \,\Delta^{(l)}\,.
\label{eq:steering_injection}
\end{equation}
An increasing value of $\lambda$ amplifies the magnitude of this vector, pushing the model more strongly toward the concept of interest  (e.g., changing the tempo or the timbre), while lower values of $\lambda$ result in subtler shifts. This method enables fine-grained control over the model’s output without requiring additional training or fine-tuning.

\subsection{Choosing the Right Layer and Injection Strategy}
\label{sec:injection_strategies}

Since \diffmean{} vectors are computed separately for each transformer layer, we obtain a set of $L$ candidate vectors, each potentially capable of steering the model’s behavior. A key question is how to best inject these vectors during inference.
We evaluate two injection strategies, illustrated in \Cref{fig:injection_strategies}:

\begin{enumerate}
    \item \textbf{All-to-All strategy}: Inject a distinct steering vector $\Delta^{(l)}$ into its corresponding layer $l$, applying layer-specific modifications throughout the network.
    \item \textbf{One-to-All strategy}: Identify a single optimal direction $\Delta^{(l_{\text{best}})}$ at a chosen layer (as described in \Cref{sec:layer-scan}) and inject this one vector across \textit{all} layers.
\end{enumerate}

Empirically, we find that the \textit{One-to-All} strategy yields superior results, both in terms of effectively steering the model toward the desired attributes and preserving the overall quality of generated music. Thus, we adopt this approach throughout the paper.

The effectiveness of applying a single direction across multiple layers can be attributed to the structure of the transformer’s residual stream: information flows continuously through attention and MLP layers, which modify activations before writing them back into the same latent space. As a result, steering vectors derived from a specific layer often remain effective when applied at different depths in the network.

\section{Experiments}
\label{sec:experiments}

\subsection{Prompt Sets and Metrics}
\label{sec:metrics}

We employed \texttt{GPT-o1} to automatically create four distinct sets of 100 textual prompts each, denoted as $S_{\text{Fast}}, S_{\text{Slow}}, S_{\text{Bright}}, S_{\text{Dark}}$, specifically chosen to induce clear variations in tempo or brightness. We run each prompt through \musicgen{}, caching the activations at the \texttt{EOS} token for every layer to compute their mean.

\begin{figure}[t]
    \begin{subfigure}{\linewidth}
        \centering
        \includegraphics[width=\linewidth]{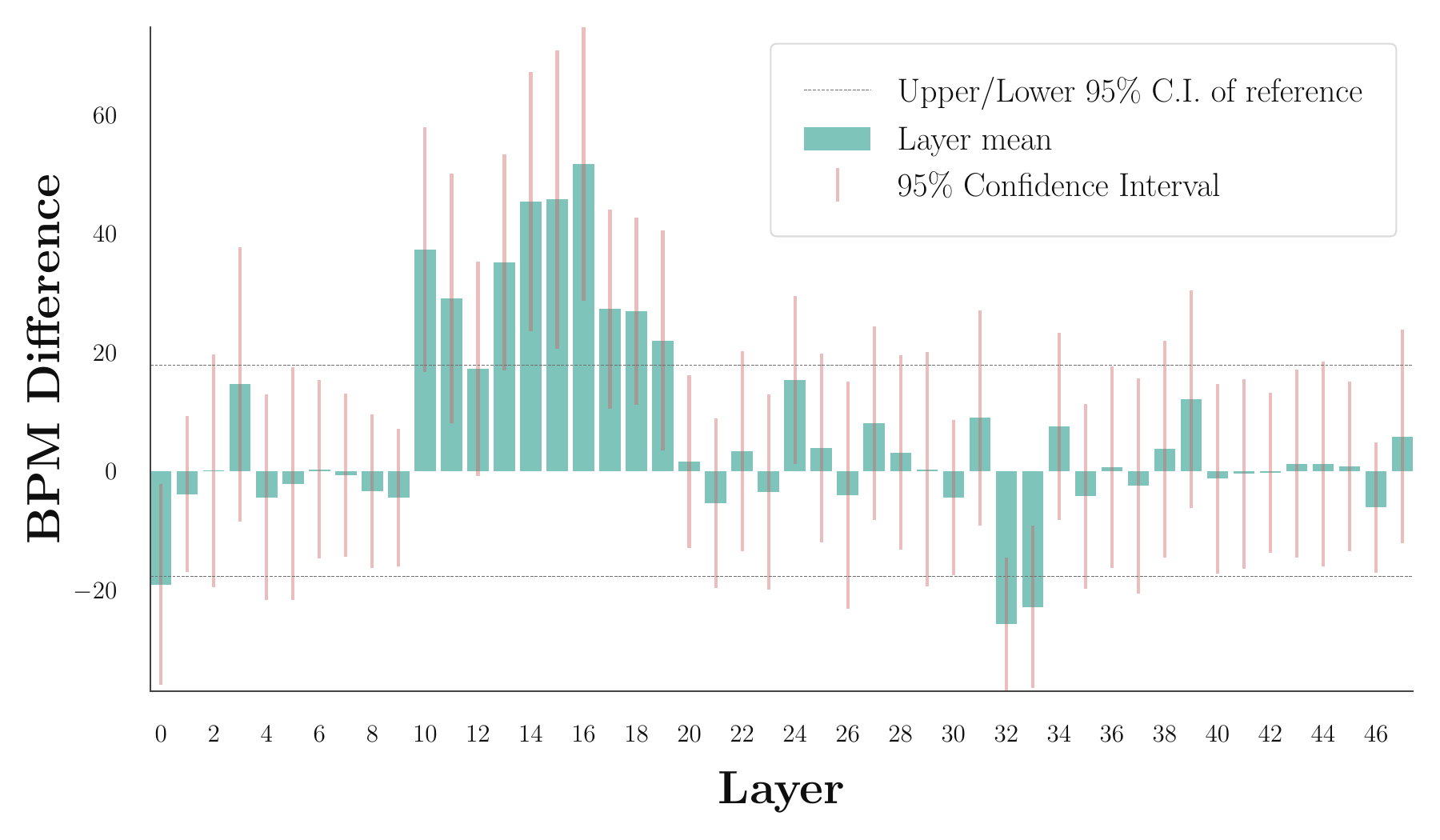}
        \caption{Speed.}
        \label{fig:tempo-steering}
    \end{subfigure}
    \hfill 
    \begin{subfigure}{\linewidth}
        \centering
        \includegraphics[width=\linewidth]{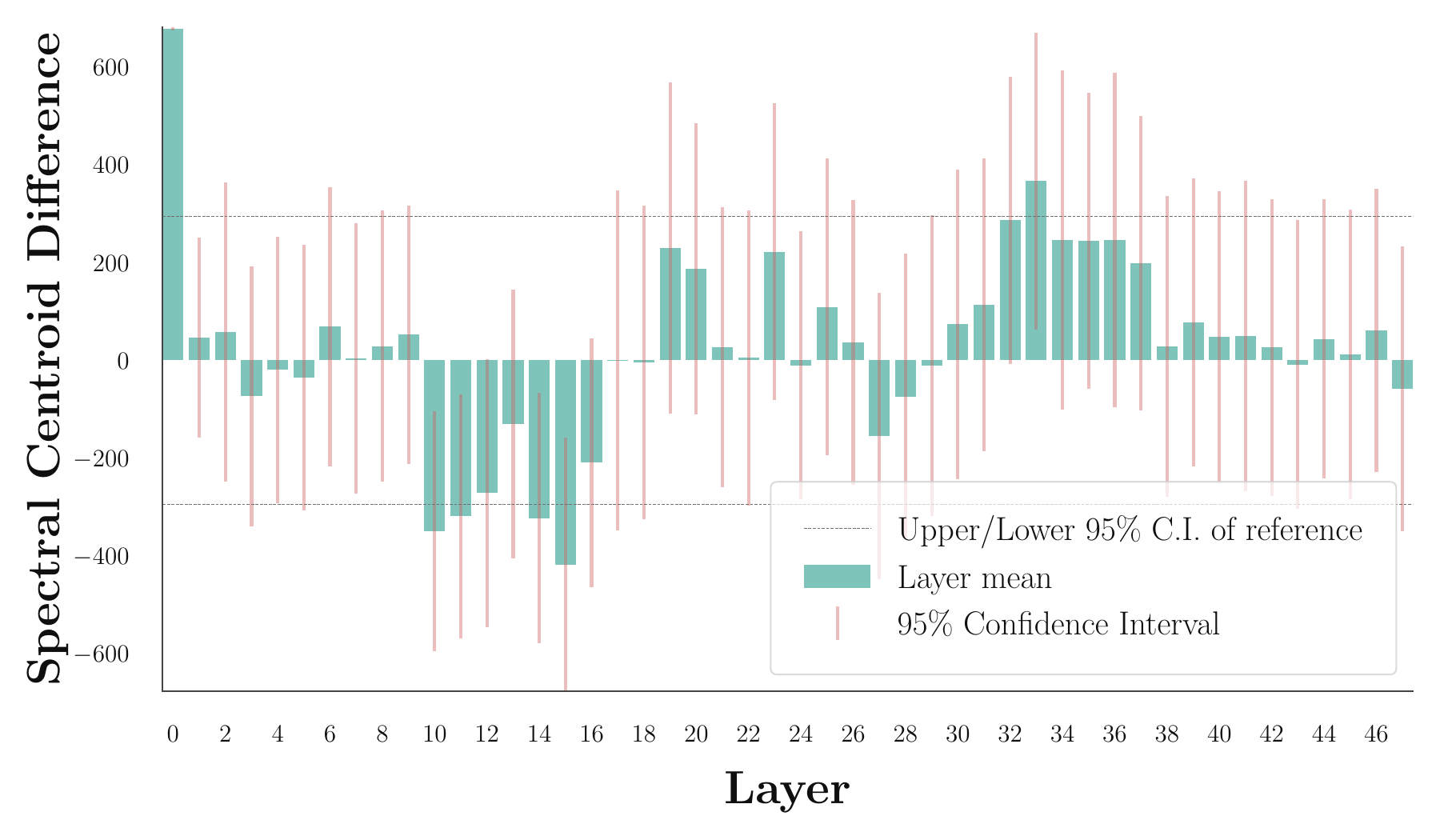}
        \caption{Timbre.}
        \label{fig:brightness-steering}
    \end{subfigure}
    \caption{Relative effect of injecting $\Delta^{(l)}$ into the baseline calculated with the original model for both speed and timbre. The mid-range block of layers 10-16 appears to be the one that induces a greater effect in both the attributes compared to the benchmark value.}
\end{figure}

To assess how effectively the steering vectors alter the model’s output, we generate music from a separate, neutral evaluation set of $n=50$ diverse prompts. We then evaluate the tempo by estimating \emph{BPM} via \texttt{BeatThis} \cite{foscarinbeat}, chosen for its greater robustness than other tested methods, and timbre by computing the \emph{Spectral Centroid}. The latter
is computed as a weighted average of the frequencies in the audio, reflecting the amount of high-frequency energy.
For faster songs, we expect a higher BPM, while music with darker timbre should exhibit a lower spectral centroid. These metrics thus serve as quantitative indicators of whether the steering vectors succeed in shifting the generated music toward faster/slower or brighter/darker outputs.

\subsection{Layer-wise Scan and Best-Layer Selection}
\label{sec:layer-scan}

\begin{figure}[t]
    \begin{subfigure}{0.48\textwidth}
        \centering
        \includegraphics[width=\linewidth]{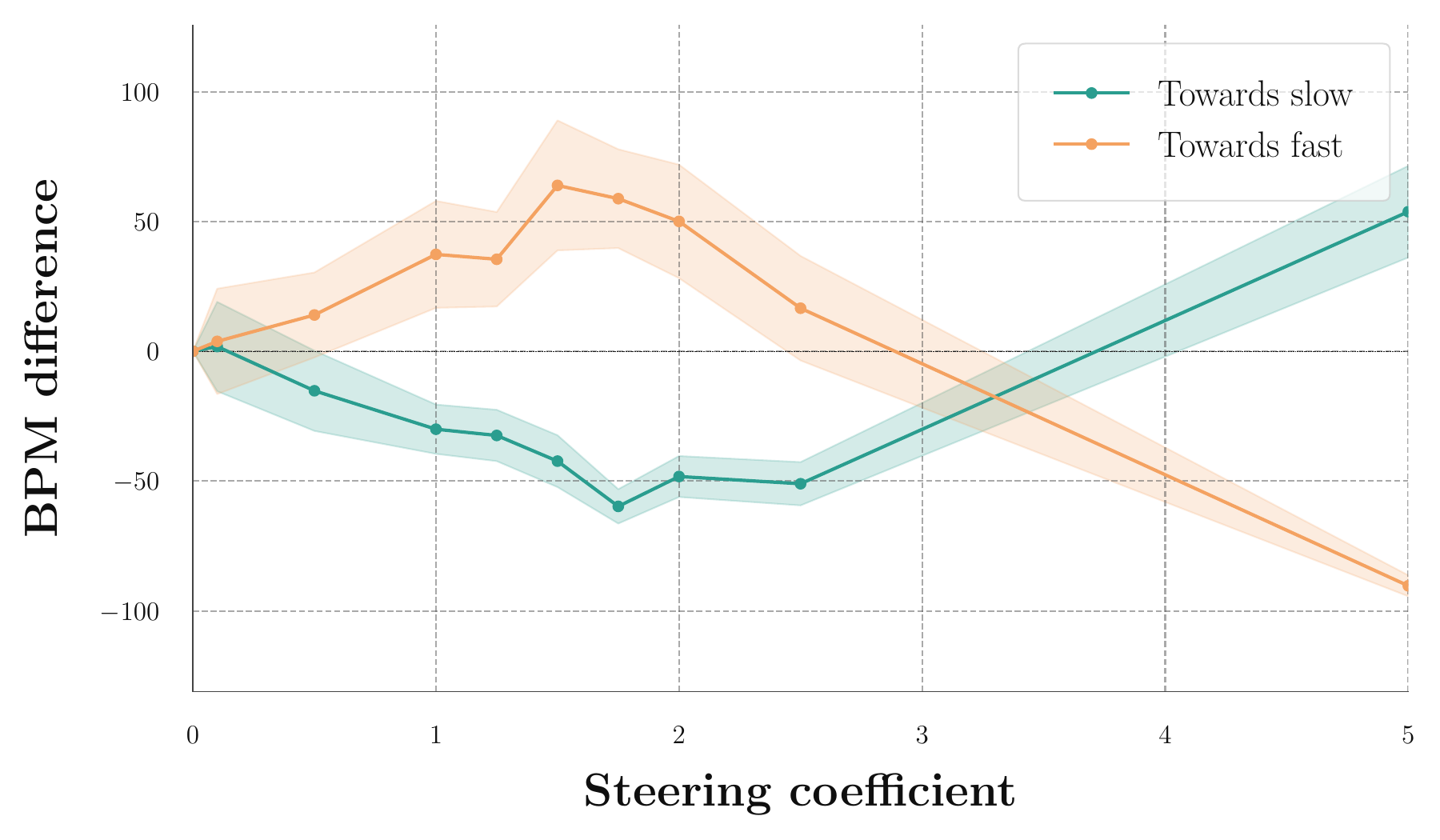}
        \caption{Tempo.}
        \label{fig:lambda-vs-bpm}
    \end{subfigure}
    \hfill 
    \begin{subfigure}{0.48\textwidth}
        \centering
        \includegraphics[width=\linewidth]{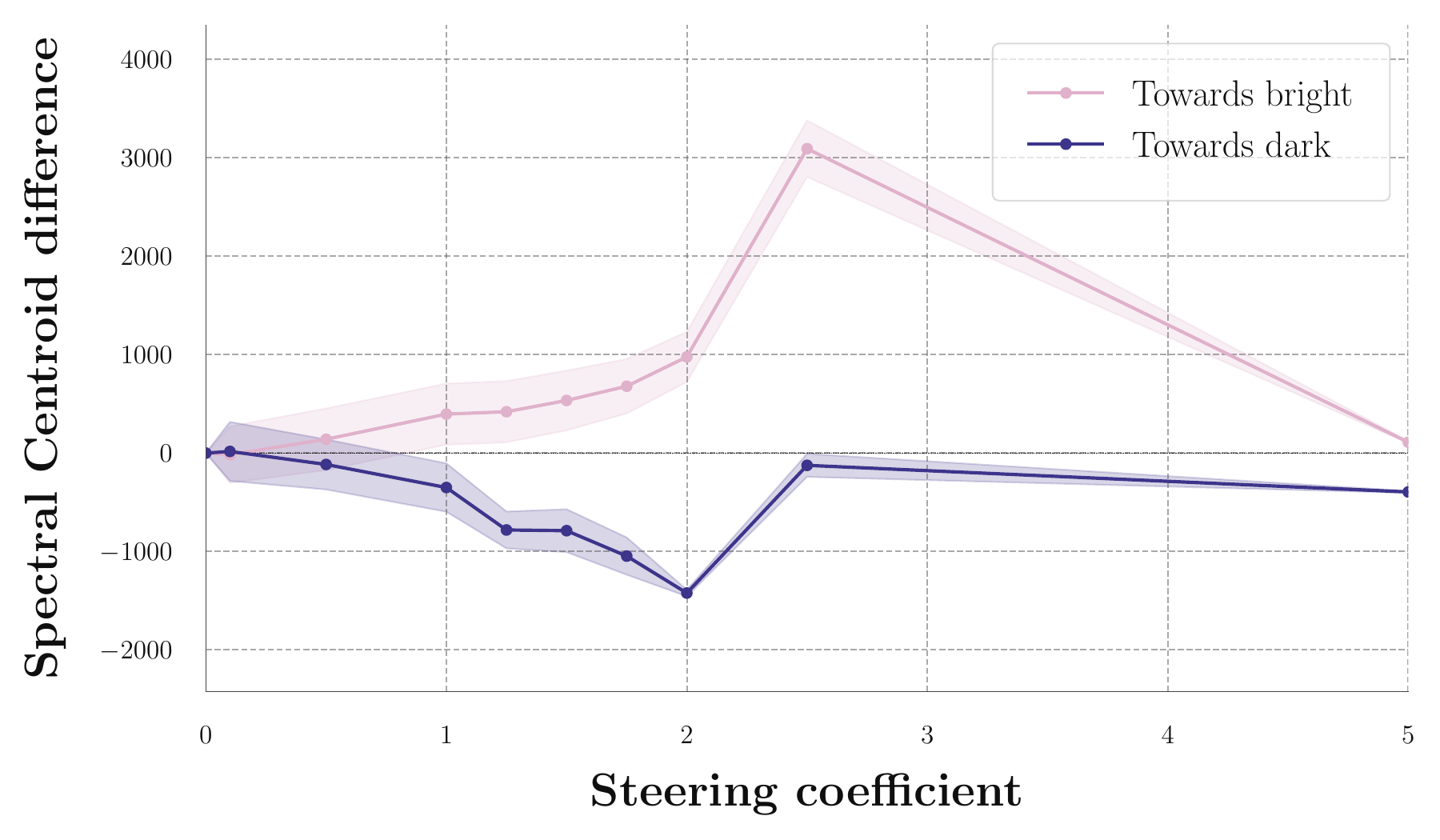}
        \caption{Timbre.}
        \label{fig:lambda-vs-centroid}
    \end{subfigure}
    \caption{Influence of the steering coefficient $\lambda$ on \{tempo, timbre\}, measured by \{BPM, spectral centroid\} (y-axis). Increasing values of $\lambda$ tend to have a greater steering effect, up to a threshold value beyond which the model reaches an out-of-distribution state. The baseline ($\lambda=0$) corresponds to no steering.}
\end{figure}

To identify the most effective layer for steering, we conduct a systematic layer-wise scan. For each transformer layer \( l = 1, \dots, L \), we compute a \diffmean{} vector \(\Delta^{(l)}\) using contrastive prompt sets (e.g., $S_{\text{Fast}}$ vs.\ $S_{\text{Slow}}$, or $S_{\text{Bright}}$ vs.\ $S_{\text{Dark}}$), as described in \Cref{sec:method}. We then evaluate each vector’s steering power by measuring how it shifts the model’s behavior when injected during inference.

To do so, we first establish baseline values for our steering metrics (e.g., BPM for tempo, Spectral Centroid for brightness) by generating music from a set of $n=50$ neutral evaluation prompts. We then iterate through each layer \(l\), injecting the corresponding vector \(\Delta^{(l)}\) during generation and re-computing the metrics on the steered outputs. The direction \(\Delta^{(l)}\) is considered more effective if it causes a consistent shift in the target metric compared to the baseline, while maintaining generation quality.

\Cref{fig:tempo-steering,fig:brightness-steering} show the results. In both cases, we observe that a mid-range block of layers—specifically layers 10--18—produces the most pronounced and consistent shifts in the desired direction. For example, \Cref{fig:tempo-steering} shows that \(\Delta^{(l)}_{\text{Fast}}\) increases BPM, while \Cref{fig:brightness-steering} shows that \(\Delta^{(l)}_{\text{Dark}}\) lowers spectral centroid, effectively darkening the timbre. These findings echo those in NLP, where mid-layer representations are known to encode higher-level semantic features \cite{skean2025layerlayeruncoveringhidden}.

\begin{figure}[t]
\centering
\includegraphics[width=0.7\columnwidth]{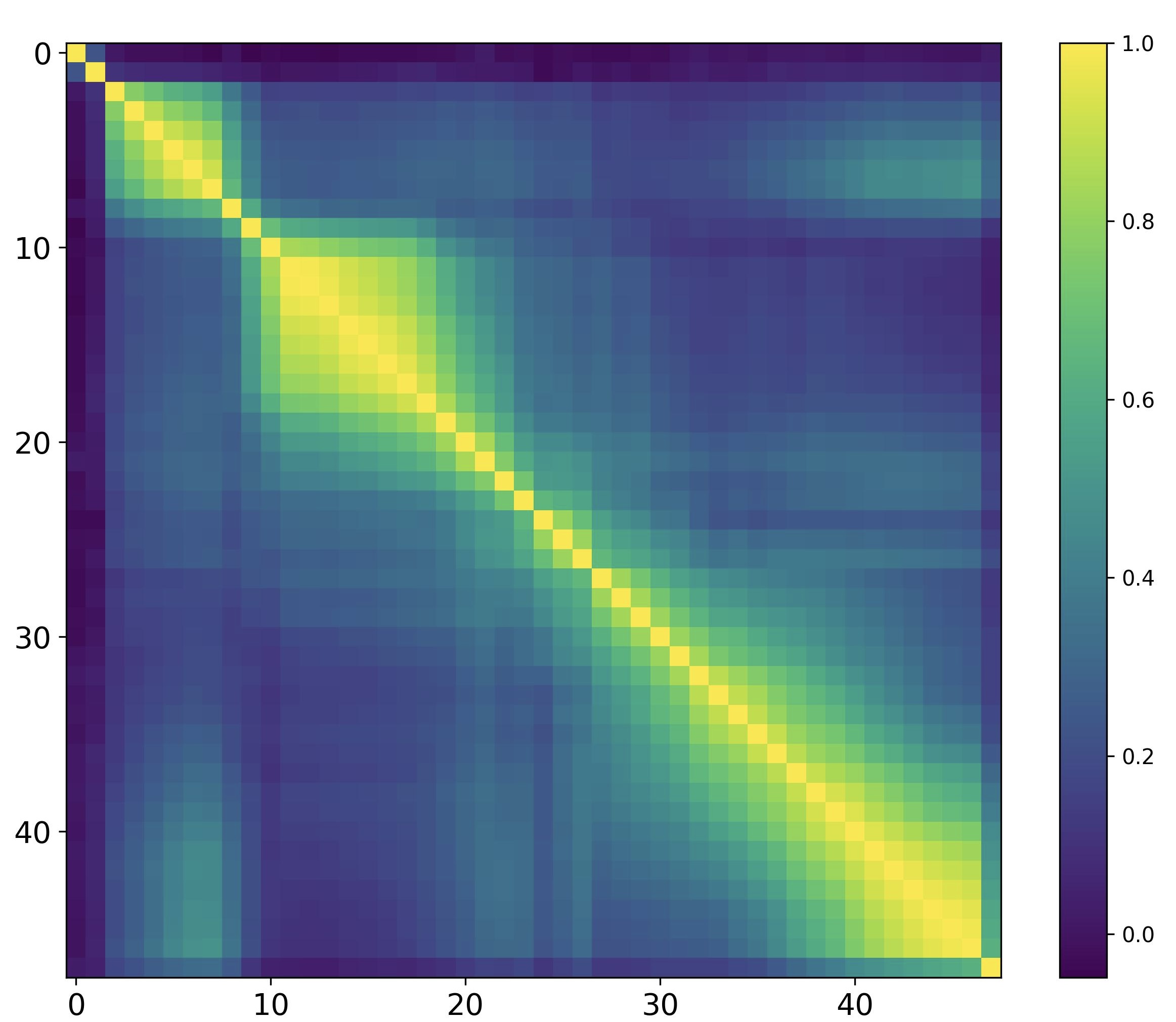}
\caption {Cosine Similarity Matrix for the directions drawn from \musicgen{}'s layers. The plot clearly shows the presence of three strongly correlated blocks. The central block of layers 10-18 is also the one that best induces the desired behavior for both Tempo and Brightness.}
\label{fig:cosine_similarity}
\end{figure}

Rather than a single ``best'' layer encoding a musical concept, our findings suggest that multiple layers act in concert to represent high-level musical attributes. \Cref{fig:cosine_similarity} further supports this view, showing that directions from layers 10--18 are strongly correlated, forming a coherent subspace. Within this cluster, we empirically choose $l_{\text{best}}$ as the layer whose \diffmean{} vector induces the strongest and cleanest steering effect in different hyperparameters configurations. We then use this direction in the One-to-All injection strategy for the rest of our experiments.

\begin{figure}
\centering
\includegraphics[width=\linewidth]{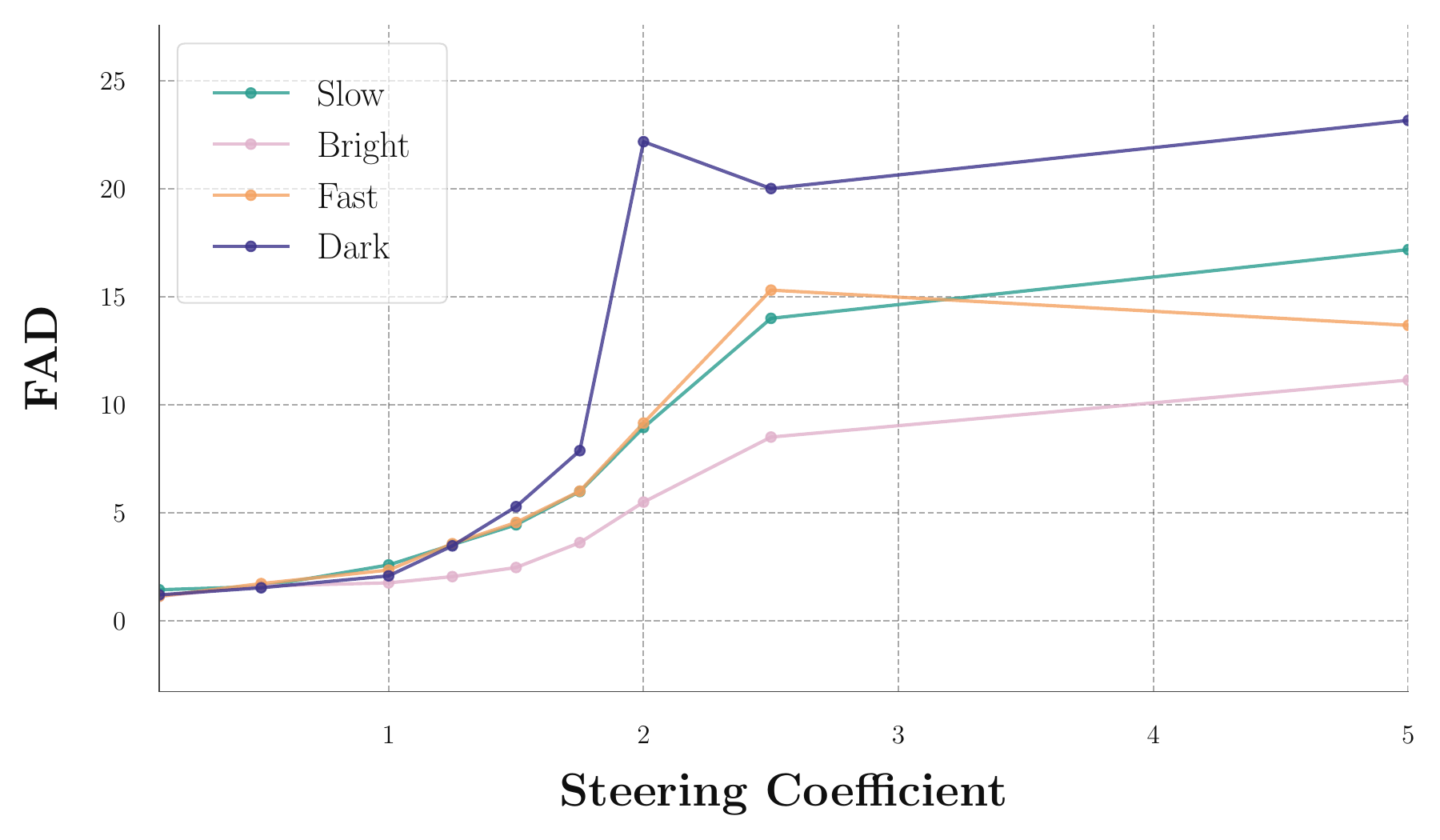}
\caption{Fréchet Audio Distance (FAD) as a function of the steering coefficient $\lambda$. For small $\lambda$, the FAD remains relatively low, indicating preserved audio quality; however, beyond $\lambda\approx 1.5$, the FAD rises sharply, suggesting that excessive steering pushes the model out of its training distribution, leading to audible artifacts.}
\label{fig:lambda-vs-fad}
\end{figure}

\subsection{Varying the Steering Coefficient}
Having identified an effective layer and steering direction, we next examine how the steering coefficient \(\lambda\) controls the intensity of the effect, varying $\lambda \in [0, 5]$. 

\Cref{fig:lambda-vs-bpm} plots BPM as a function of \(\lambda\) for both slow and fast steering directions, while \Cref{fig:lambda-vs-centroid} does the same for spectral centroid in bright vs.\ dark steering. For \(\lambda=0\), no steering is applied and the model defaults to its baseline behavior. Increasing \(\lambda\) leads to stronger shifts in the target metric, for both tempo and brightness attributes.

However, this behavior becomes inconsistent for larger values of $\lambda$. As shown in \Cref{fig:lambda-vs-fad}, the Fréchet Audio Distance (FAD) begins to rise sharply beyond $\lambda \approx 1.5$, indicating a loss in audio quality. At this point, steering becomes unstable: generated music may exhibit artifacts and diverge from the intended concept. Within a moderate range (roughly \(\lambda \in [0,1.5]\)), we observe smooth, predictable shifts in tempo or brightness that align well with the target concept.

These results suggest that interpretability-based steering can effectively modulate musical attributes when applied with a moderate \(\lambda\). Lower values deliver clear shifts in tempo or timbre while preserving realism, whereas extreme values risk driving the model beyond its typical operating regime.

\subsection{Impact of the Prompt Dataset Size on Steering}
\begin{figure}[t]
\centering
\includegraphics[width=\linewidth]{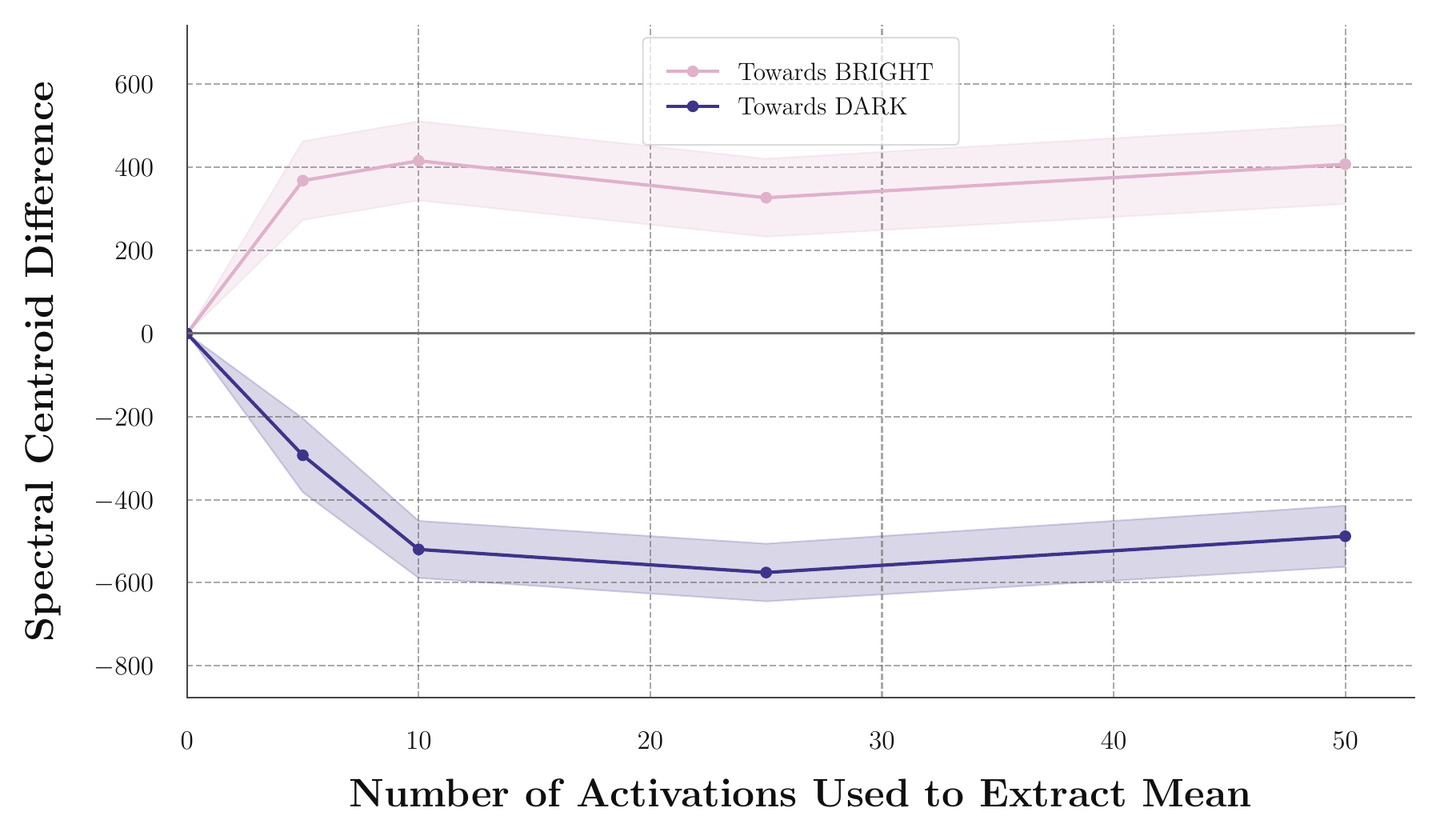}
\caption{Influence of dataset size on timbre steering. The x-axis represents the number of prompts used to compute the \diffmean{} vectors for ``bright'' and ``dark,'' while the y-axis shows the resulting spectral centroid difference of the steered outputs. Even 10 prompts suffice to induce a significant shift, allowing quick, ad hoc construction of effective steering sets. Additional prompts yield more stability but show diminishing returns.}
\label{fig:n_prompts_vs_centroid}
\end{figure}

\Cref{fig:n_prompts_vs_centroid} illustrates how the spectral centroid (y-axis) of the steered outputs varies as a function of the number of \textit{prompts} (x-axis) used to compute the \diffmean{} vectors for ``bright'' and ``dark'' timbre. Notably, even with as few as 10 prompts, we observe a clear shift in brightness, suggesting that an effective steering direction can be established quickly. Adding more prompts refines this direction and improves consistency, but the largest gains occur in the first 10--25 samples. Beyond that, the effect plateaus, implying that a relatively small dataset can already provide robust control over timbre.

\subsection{Overall Steering Results}
\label{sec:results}
To test generalization, we created a held-out set of $n = 50$ prompts, structurally similar to the evaluation set. Based on the results shown in \Cref{fig:tempo-steering} and \Cref{fig:brightness-steering}, we selected the layers that appeared to exhibit the most effective behavior—layer~16 for Tempo and layer~10 for Timbre. Furthermore, as shown in \Cref{fig:lambda-vs-fad}, we chose a value of $\lambda = 1.25$, which offered a good trade-off between effectiveness on the target attributes and the FAD score.

\Cref{tab:table-experiments} reports the results obtained on this new set. Across the four attributes, we observe a relative improvement ranging from 20\% to 40\%. For Tempo, this translates to a shift of approximately $30$-$50$ BPM, while for Timbre, the Spectral Centroid increases by 360 to 720 units. 

\renewcommand{\arraystretch}{1.4}

\begin{table}[h]
    \centering
    \resizebox{1\linewidth}{!}{%
        \begin{tabular}{c c c c c}
            \toprule
            Attribute & Variant     & Absolute     & Relative ($\uparrow$)  & $\textrm{FAD}$ ($\downarrow$) \\
            \cmidrule{1-1} \cmidrule(l){2-2} \cmidrule(l){3-5} \rowcolor{gray!15}
            \cellcolor{white} & \musicgen{} & $134.54 \pm 14.38$ & -- & -- \\ 
            & $\rightarrow$ slow & $85.92 \pm 8.48$  & $36.1\%$ & $6.76$ \\ 
            \cellcolor{white} \multirow{-3}{*}{\small \vertical{Tempo}}  & $\rightarrow$ fast &  $176.02 \pm 17.35 $ & $30.1\%$ & $4.64$ \\ 
            \midrule \rowcolor{gray!15}
            \cellcolor{white} & \musicgen{} & $1799.42 \pm 270.22$ & -- & -- \\
            & $\rightarrow$ bright & $2161.15 \pm 303.72$ & $20\%$ & $1.78$ \\ 
            \cellcolor{white} \multirow{-3}{*}{\small \vertical{Timbre}} & $\rightarrow$ dark & $ 1076.80 \pm 200.04$ & $40.1\%$ & $3.49$ \\
            \bottomrule
        \end{tabular}
    }
    \caption{Steering effectiveness with $\lambda=1.25$. Tempo is measured in BPM, while timbre is measured by the spectral centroid introduced in \cref{sec:metrics}.}
    \label{tab:table-experiments}
\end{table}

\section{Discussion}
\label{sec:discussion}

In this section, we first discuss how steering may enable a more granular form of control when compared to prompting, and then how the framework can be extended to different pairs of attributes.

\subsection{Steering for More Granular Control}
While prompt engineering is a common method for guiding music generation models, it offers limited precision. Changing the prompt simply re-samples from the model’s distribution, rather than providing true control or targeted editing. A more effective alternative is to intervene directly in the model’s internal activations. By identifying and manipulating specific latent directions, we can achieve finer-grained adjustments and gain deeper insight into the model’s behavior. This approach also opens up practical applications, such as real-time, nuanced control in digital audio production via specialized plug-ins.

\subsection{Extending to Other Attribute Pairs}
While our experiments focused on two foundational musical attributes, namely tempo (fast vs. slow) and timbre (bright vs. dark), the methodology is inherently general and can be applied to any pair of semantically opposite attributes. In principle, this merely requires defining two sets of contrastive prompts representative of the target attributes (e.g., ``loud'' vs. ``soft'', ``major'' vs. ``minor'', or any other binary dimension) and then using the same difference-of-means approach to derive a steering direction from their mean activations. While a strict metric may be harder to define for more subtle or subjective attributes (e.g., ``energetic'' vs. ``calm'', ``complex'' vs. ``minimalist''), the evaluation procedure can rely on a simple classifier to distinguish which side of the attribute each generation tends toward. By tuning a small binary model on relevant reference data, one can automatically label generated samples during inference and measure the effectiveness of steering.

\section{Conclusions}
\label{sec:conclusions}

We presented a novel methodology for interpreting and controlling musical attributes in large text-to-music models through \textit{activation-based steering}. Our experiments focused on two fundamental binary attributes—tempo and timbre—and showed that the \diffmean{} method successfully identifies latent directions that capture these concepts within \musicgen{}’s residual stream. Injecting a single steering vector at all layers allowed for fine-grained, continuous control over the target attribute, without requiring additional training or fine-tuning. Doing so, we discovered that mid-layer activations (particularly layers 10--18) exhibit a concentrated capacity for modulating tempo and brightness, suggesting a structured representation of higher-level musical features. Varying the steering coefficient \(\lambda\) showed that moderate values produce smooth, predictable shifts, while extreme values lead to out-of-distribution outputs and degraded audio quality.

Beyond showcasing an effective way to steer music generation, these findings shed light on the internal mechanisms of large audio models while opening avenues for future work in \textit{mechanistic interpretability} applied to multimodal generative systems. Directions for improvement include expanding to more nuanced musical attributes (e.g., emotion, style), combining multiple steering vectors to achieve mixed concepts (e.g., “fast but dark”), and investigating how domain constraints—such as tonal theory—could be integrated into the interpretability pipeline. More broadly, this approach provides a promising foundation for understanding how large models encode rich, non-linguistic information, ultimately advancing transparency and control in generative AI.


\bibliography{references/mech_int, references/audio, references/misc}

\clearpage

\appendix

\section{Implementation details}

\subsection{Injection Across Autoregressive Steps}
There are multiple possibilities for injecting a specific direction $\Delta^{(l)}$ into the intermediate activations of a text-to-music model. In \Cref{sec:method}, we have already discussed the choice of extracting this vector from the hidden state corresponding to the EOS token, as well as the injection strategies outlined in \Cref{sec:injection_strategies}. However, additional considerations must be addressed. Given that \musicgen{} is an autoregressive model, a primary consideration is whether to inject the vector $\Delta^{(l)}$ only during the first autoregressive step or consistently throughout all steps. We selected the latter approach, as it provided superior results and enhanced the quality of generated music.

\subsection{Injection into CFG Branches}
\musicgen{} employs classifier-free guidance (CFG) to steer the music generation process. An important implementation decision concerns the CFG branches into which $\Delta^{(l)}$ should be injected. Specifically, while injecting exclusively into either the conditional or unconditional CFG channel is feasible, we opted for injecting into both branches, as this strategy consistently yielded improved generation quality and stability.

\subsection{Tempo Prompts} 
We provide here examples of tempo-related prompts. 
Examples of slow prompts may be 
\ex. ``\emph{Calm acoustic guitar tune with a gentle, soothing vibe}''

\ex. ``\emph{A restful choir piece in a cathedral-like setting}''

while fast prompts could be

\ex. ``\emph{Energic acoustic guitar tune with a lively, uplifting vibe}''

\ex. ``\emph{An up-tempo R\&B track with an exhilarating vocal line}''.

\subsection{Timbre prompts}
We provide here examples of timbre-related prompts. Brightness-inducing prompts are for example 
\ex. ``\emph{A vibrant and brilliant orchestral fantasy theme}''

\ex. ``\emph{A crisp and sparkling electronic dance track}''

while prompts that could induce darker tunes could be

\ex. ``\emph{A deep and subdued orchestral fantasy theme}''

\ex. ``\emph{A somber and muted electronic dance track}''.

\section{Additional experiments}

\subsection{Visualizing Frequency Shifts Due to Steering Vector Injections}
\begin{figure}[htbp]
  \centering
  \begin{subfigure}{\columnwidth}
    \centering
    \includegraphics[width=\columnwidth]{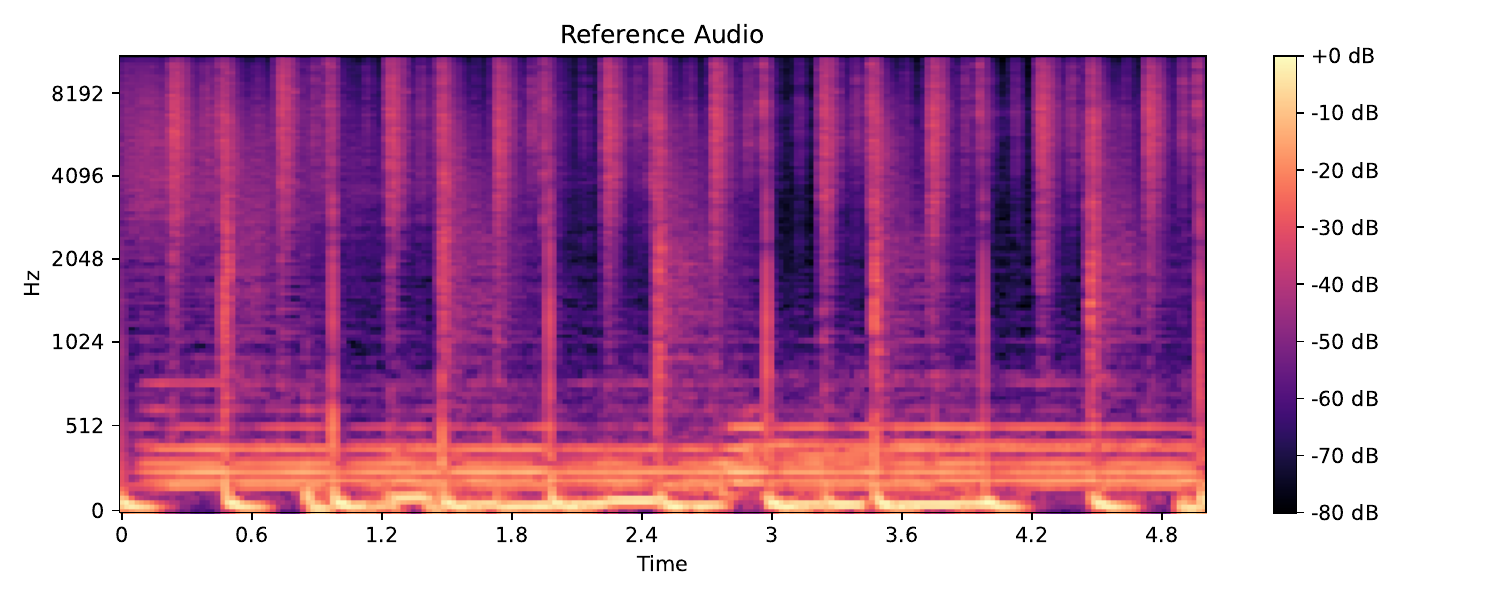}
    \caption{Reference Spectrogram}
    \label{fig:reference_spectrogram}
  \end{subfigure}
  
  \vspace{0.4cm}
  
  \begin{subfigure}{\columnwidth}
    \centering
    \includegraphics[width=\columnwidth]{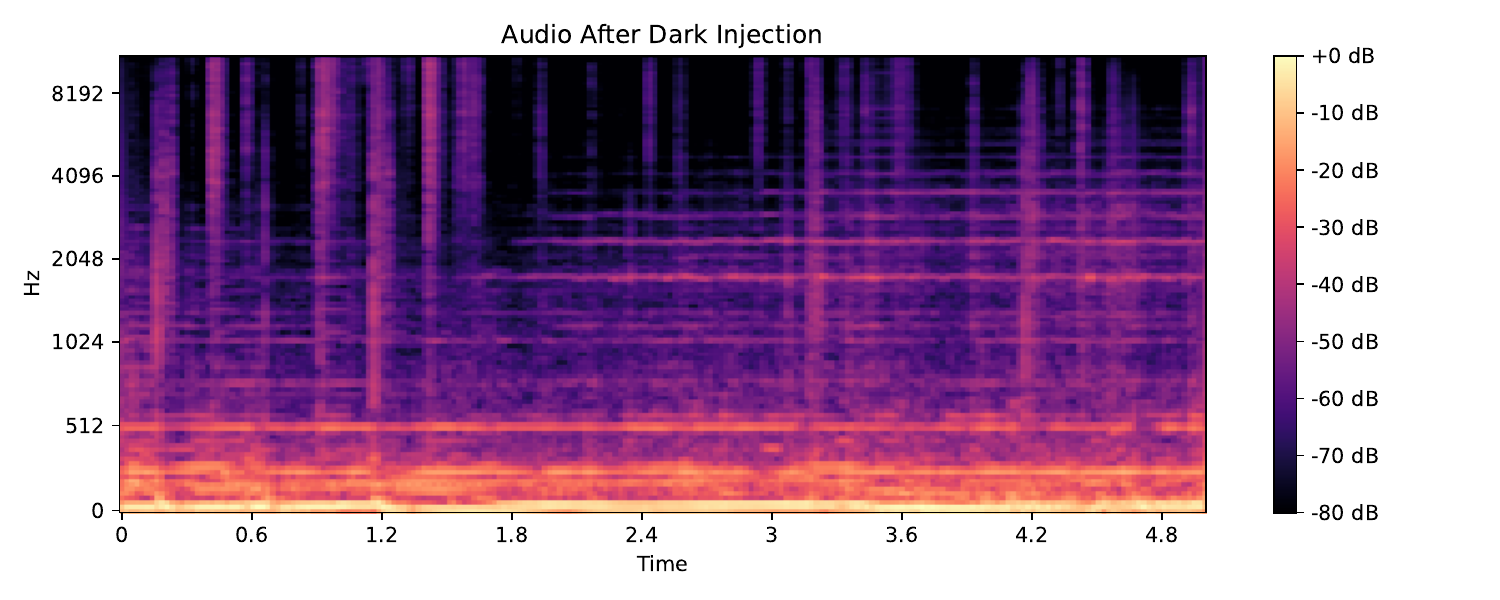}
    \caption{Spectrogram After Dark Injection}
    \label{fig:dark_spectrogram}
  \end{subfigure}
  
  \vspace{0.4cm}
  
  \begin{subfigure}{\columnwidth}
    \centering
    \includegraphics[width=\columnwidth]{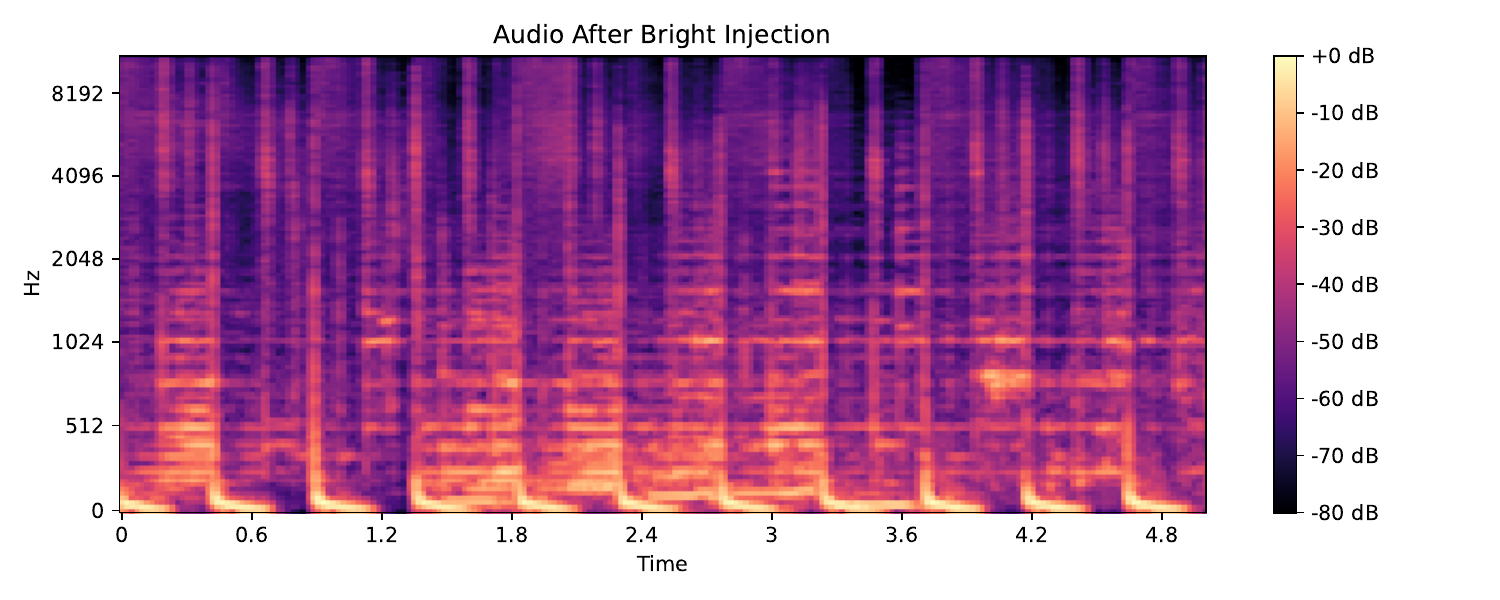}
    \caption{Spectrogram After Bright Injection}
    \label{fig:spectrogram_bright}
  \end{subfigure}
  
  \caption{Comparison of spectrograms under different injection conditions: (a) the reference audio, (b) audio after the dark injection, and (c) audio after the bright injection, highlighting the corresponding spectral shifts.}
  \label{fig:spectrograms}
\end{figure}

We provide a detailed visualization of the spectral changes induced by the injection of steering vectors aimed at modulating timbre. As shown in \Cref{fig:reference_spectrogram}, the reference audio exhibits a balanced frequency distribution. In contrast, \Cref{fig:dark_spectrogram} illustrates that injecting a dark-oriented vector systematically attenuates higher frequencies, thereby shifting the timbre toward a darker sound. Conversely, \Cref{fig:spectrogram_bright} demonstrates that a bright-oriented injection amplifies the high-frequency components, resulting in a perceptually brighter audio signal. These figures collectively confirm that our method produces the expected directional changes in the frequency spectrum.

\end{document}